\documentclass{elsart4}

\usepackage{amssymb}

\usepackage{graphics}
\usepackage{graphicx}
\usepackage{epsfig}
\usepackage{amssymb}
\journal{Physica B}

\begin{document}

\begin{frontmatter}


 \title{D-wave-like nodal superconductivity in the organic conductor (TMTSF)$_2$ClO$_4$} 
 \author{A.G. Lebed\corauthref{cor1}}
 \address{Department of Physics, University of Arizona, 1118 E. 4th Street, Tucson, Arizona 85721, USA}
 \address{L.D. Landau Institute for Theoretical Physics, 2 Kosygina Street, 117334 Moscow, Russia}
 \corauth[cor1]{Tel: +1 520 626 1031; FAX:+1 520 621 4721; e-mail: lebed@physics.arizona.edu}





\begin{abstract}
We suggest theoretical explanation of the high upper critical magnetic
field, perpendicular to conducting chains, $H^{b'}_{c2}$, experimentally observed 
in the superconductor (TMTSF)$_2$ClO$_4$, in terms of singlet superconducting 
pairing.
In particular, we compare the results of d-wave-like nodal, d-wave-like node-less, and 
s-wave scenarios of superconductivity.
We show that, in d-wave-like nodal scenario, superconductivity can naturally exceed 
both the orbital upper critical magnetic field and Clogston-Shandrasekhar  paramagnetic limit as well as reach experimental value, $H^{b'}_{c2} \simeq 6 \ T$, in contrast to d-wave-like 
node-less and s-wave 
scenarios.
In our opinion, the obtained results are strongly in favor of d-wave-like nodal superconductivity
in (TMTSF)$_2$ClO$_4$, whereas, in a sister compound, (TMTSF)$_2$PF$_6$,
we expect either the existence of triplet order parameter or the coexistence of triplet and 
singlet order parameters. 
\end{abstract}

\end{frontmatter}

\section{Introduction}
\label{1}
High magnetic field properties of (TMTSF)$_2$X (X=ClO$_4$, PF$_6$, 
etc.) organic materials have been intensively studied [1] since the discovery of superconductivity in (TMTSF)$_2$PF$_6$ conductor [2].
From the beginning, it was clear that superconductivity in the above mentioned
materials was unconventional.
Indeed, early experiments [3,4] showed that the Hebel-Slichter peak was absent in
the NNR measurements [3] and that superconductivity is destroyed by non-magnetic
impurities [4].
These facts were strong arguments that superconducting order parameter 
changed its sign on the quasi-one-dimensional (Q1D) Fermi surfaces (FS) of 
(TMTSF)$_2$X material.
The main results of both experiments [3,4] were recently confirmed in a number 
of publications (see, for example, [5,6]).
It is important that the above mentioned  experiments did not contain information 
about spin part of superconducting order parameter and could not distinguish 
between singlet and triplet superconducting pairings.

The first measurements of the Knight shift in (TMTSF)$_2$PF$_6$ conductor [6,7]
showed that it was not changed in superconducting phase, which was interpreted
in favor of triplet superconductivity [6,7].
On the other hand, the more recent Knight shift measurements [8] in superconductor (TMTSF)$_2$ClO$_4$ have shown a clear change of the Knight shift in
superconducting phase at relatively low magnetic fields, $H \simeq 1 \ T$, and have
been interpreted as evidence of singlet superconductivity [8].
Another argument in favor of singlet order parameter is the fact that the upper
critical magnetic field, parallel to conducting axis, $H^a_{c2}$ [9], is paramagnetically limited [10].
Moreover, very recently the Larkin-Ovchinnikov-Fulde-Ferrell phase [11,12], which
appears for singlet superconducting pairing, has been experimentally discovered 
[13,14] in (TMTSF)$_2$ClO$_4$ and theoretically interpreted [15].

\section{Goal}
\label{2}
In such situation, where support of singlet superconducting pairing in (TMTSF)$_2$ClO$_4$ material is increasing, it is important to reinvestigate theoretically
high experimental upper critical magnetic fields, $H^{b'}_{c2}$ [13,14,16], observed 
for a magnetic field, perpendicular to conducting chains.
For many years, large values of $H^{b'}_{c2}$ have been considered as a consequence
of triplet superconducting pairing.
Our goal is to show that we can naturally explain large values of $H^{b'}_{c2}$ 
within singlet d-wave-like nodal scenario of superconductivity in 
(TMTSF)$_2$ClO$_4$.
We also show that d-wave nod-less and s-wave scenarios are much less consistent 
with experimental value of $H^{b'}_{c2} \simeq 6 \ T$ 
[13,14].
This value exceeds both the quasi-classical upper critical field [17] and Clogston-Shandrasekhar paramagnetic limit [18] due to the coexistence of two unusual
superconducting phases: Reentrant superconductivity [19-24] and Larkin-Ovchinnikov-Fulde-Ferrell phase [11,12].

\section{Results}
\label{3}

Let us consider Q1D spectrum of (TMTSF)$_2$ClO$_4$ conductor in tight binding
model [1],
\begin{equation}
\epsilon({\bf p})= - 2t_a \cos(p_x a/2) - 2 t_b \cos(p_y b') - 2t_c
\cos (p_z c^*),
\end{equation}
in a magnetic field, perpendicular to its conducting chains,
\begin{equation}
{\bf H} = (0,H,0), \ \ {\bf A} = (0,0,-Hx),
\end{equation}
where $t_a \gg t_b \gg t_c$ are electron hoping integrals along ${\bf a}$ , ${\bf b'}$, 
and ${\bf c^*}$ axes, respectively.
Electron spectrum (1) can be linearized near two slightly corrugated sheets of 
Q1D FS as
\begin{equation}
\delta \epsilon^{\pm}({\bf p})= \pm v_x(p_y)[p_x \mp p_F(p_y)] - 2t_c \cos(p_z c^*) ,
\end{equation}
where +(-) stands for right (left) sheet of Q1D FS.

We represent electron wave functions in a real space in the following way:
\begin{eqnarray}
\Psi_{\epsilon}^{\pm}(x,y,z,\sigma) = \exp[i p_F(p_y)x] &&\exp(ip_y y) \exp(ip_z z) 
\nonumber\\
&&\times \psi_{\epsilon}^{\pm} (x,p_y,p_z,\sigma).
\end{eqnarray}
Let us use the Peierls substitution method,
\begin{equation}
p_x \mp p_F(p_y) \rightarrow -i d/dx, \ \ p_z \rightarrow p_z - eA_z/c.
\end{equation}
As a result, the Schrodinger-like equation for wave functions $\psi^{\pm}(x,p_y,p_z,\sigma)$
can be written as
\begin{eqnarray}
\biggl[ \mp i v_x(p_y) \frac{d}{dx} &&- 2t_c \cos \biggl(p_zc^*+\frac{\omega_c}{v_F}x \biggl)
- \mu_B \sigma H \biggl] \nonumber\\
&&\times \psi_{\epsilon}^{\pm}(x,p_y,p_z,\sigma) =
 \delta \epsilon \ \psi_{\epsilon}^{\pm}(p_x,p_y,p_z,\sigma),
\end{eqnarray}
where $\mu_B$ is the Bohr magneton, $\sigma=\pm1$ stands for spin up 
and down, respectively; $\omega_c = eHv_F c^*/c$, 
$\delta \epsilon = \epsilon - \epsilon_F$.

It is important that Eq.(6) can be analytically solved:
\begin{eqnarray} 
&&\psi_{\epsilon}^{\pm}(x,p_y,p_z,\sigma)=
\frac{\exp[\pm i \delta \epsilon x /v_x(p_y)]}{\sqrt{2 \pi v_x(p_y)}}
\exp \biggl[\pm i \frac{\mu_B \sigma H x}{v_x(p_y)} \biggl]
\nonumber\\
&&\times \exp \biggl[ \pm i \frac{2t_c}{v_x(p_y)} \int_0^{x} \cos \biggl( p_z c^*
+ \frac{\omega_c}{v_F}u \biggl) du \biggl] .
\end{eqnarray}
The corresponding Green functions can be obtained from the following equation
(see Ref.[25]):
\begin{equation}
G(x,x_1,p_y,p_z,\sigma)= \sum_{\epsilon} \frac{\psi^*_{\epsilon} (x,p_y,p_z,\sigma)
\psi_{\epsilon} (x_1,p_y,p_z,\sigma)}{i \omega_n - \epsilon} .
\end{equation}

Below, we introduce superconducting order parameter in the following way:
\begin{equation}
\Delta(p_y,x) = f(p_y b') \Delta(x), \ \ \int^{2 \pi}_0  f^2(p_y b') d(p_y b')/2 \pi= 1,
\end{equation}
where function $f(p_yb')$ takes into account three possible order parameters: 
d-wave-like nodal, $f(p_yb') = \sqrt{2} \cos(p_yb')$, s-wave, $f(p_yb')=1$, and 
d-wave-like node-less, $f(p_yb') = 2 \theta(p_y b'+\pi/2) - 2 \theta(p_y b' -\pi/2) -1$ 
[26].
Linearized gap equation for all three possible singlet scenarios of superconductivity
in (TMTSF)$_2$ClO$4$ conductor can be derived, using the Gor'kov equations
[25] for non-uniform superconductivity [28-30].
As a result of rather lengthy but straightforward calculations, we obtain:
\begin{eqnarray}
&&\Delta(x) = \tilde g  \int \frac{d p_y}{v_x(p_y)} \int_{|x-x_1| > \frac{v_x(p_y)}{\Omega}} 
\frac{2 \pi T dx_1}{v_x(p_y) \sinh[ \frac{ 2 \pi T |x-x_1|}{v_x(p_y)}]}
\nonumber\\
&&\times J_0 \biggl\{ \frac{8 t_c v_F}{\omega_c v_x(p_y)} 
\sin \biggl[ \frac{\omega_c (x-x_1)}{2v_F} \bigg]
\sin \biggl[ \frac{\omega_c (x+x_1)}{2v_F} \bigg] 
\biggl\} 
\nonumber\\
&&\times \cos \biggl[ \frac{2 \beta \mu_B H (x-x_1)}{v_x(p_y)} \biggl] \
f^2(p_yb') \Delta (x_1) ,
\end{eqnarray}
where $\tilde g$ is effective electron coupling constant, $\Omega$ is cut-off
energy, parameter $\beta$ takes into account possible deviations of superconductivity
in  (TMTSF)$_2$ClO$4$ from weak coupling scenario.

Note that Eq.(10), derived in this article, is very general. It takes into account 
both the orbital and paramagnetic effects against 
superconductivity. 
In particular, it takes into account quantum nature of electron motion along open 
orbits in the extended  Brillouin zone in Q1D conductor (3) in a magnetic 
field (2).
It is possible to make sure that the main quantum parameter in Eq.(10) is
$2 t_c v_F/\omega_c v_x(p_y) \simeq 2t_c/\omega_c$.
Let us estimate the value of this quantum parameter, using quasi-classical
language. 
In accordance with Ref.[19], quasi-classical electron trajectory in a magnetic
field (2) can be written as
\begin{equation}
z(t,H) =  c^* \ l_{\perp}(H)  \cos(\omega_c t) ,
\end{equation}
where $l_{\perp}(H) =2t_c/\omega_c$ corresponds to a "size" of electron trajectory 
in terms of interlayer distance, $c^*$; $t$ is time.

It is possible to show that 
\begin{equation}
l_{\perp}(H) = \frac{2 \sqrt{2}}{\pi} \frac{\phi_0}{a c^* H} \frac{t_c}{t_a} \simeq
\frac{2 \times 10^3}{H(T)} \frac{t_c}{t_b} \frac{t_b}{t_a} ,
\end{equation}
where $H(T)$ is a magnetic field, measured in Teslas, $\phi_0$ is the flux
quantum.
Note that value of $t_a/t_b \simeq10$ is very well known in (TMTSF)$_2$ClO$_4$ 
from theoretical fitting [31] of the so-called Lee-Naughton-Lebed
oscillations [31,32].
As to ratio $t_c/t_b$, it can be evaluated from the measured Ginzburg-Landau
(GL) slopes of the upper critical magnetic fields in (TMTSF)$_2$ClO$_4$ conductor 
[13,14]:
\begin{equation}
t_c/t_b= (b^*/\sqrt{2} c^*) (H^c_{c2}/H^{b'}_{c2})_{GL}  ,
\end{equation}
\begin{equation}
t_c/t_b= (b^*/ c^*) (H^c_{c2}/H^{b'}_{c2})_{GL} ,
\end{equation}
where Eq.(13) is valid for d-wave-like nodal pairing, whereas Eq.(14) is valid for
both d-wave-like node-less and s-wave pairings.

As a result, we obtain 
\begin{equation}
l_{\perp}(H=6 \ T) \simeq 0.48
\end{equation}
for d-wave-like nodal scenario of superconductivity and
\begin{equation}
l_{\perp}(H=6 \ T) \simeq 0.68 
\end{equation}
for d-wave-like node-less and s-wave ones.
Low values of the parameter $l_{\perp}(H=6 \ T)$ show that both cases
correspond to $3D \rightarrow 2D$ dimensional crossover of electron motion [19], 
where electrons are almost localized on conducting 
layers.
In this situation, superconductivity becomes almost two-dimensional and we
can approximate the Bessel function in Eq.(10) as 
$J_0(z) \simeq z^2/4$.

Below we consider the gap equation (10) at zero temperature, $T=0$.
In this case and under condition of $3D \rightarrow 2D$ dimensional crossover, 
it is possible to represent superconducting order parameter in the following 
way:
\begin{eqnarray}
\Delta (x) = \exp(ikx)[1 &&+ \alpha_1 \cos(2 \omega_c x /v_F)
\nonumber\\
&&+ \alpha_2 \sin(2 \omega_c x /v_F) ],
\end{eqnarray}
where $|\alpha_1|, |\alpha_2| \ll 1$.
For such order parameter Eq.(10) can be rewritten as
\begin{eqnarray}
\frac{1}{g} = \int^{2 \pi /b'}_{0} \frac{d( p_y b')}{2 \pi }  \int^{\infty}_{\frac{v_F}\Omega} \frac{d z}{z}
f^2(p_y b') \cos \biggl( \frac{2 \beta \mu_b H z}{v_F} \biggl)
\nonumber\\
\times  \frac{v_F}{v_x(p_y)}  \biggl[1-2 l^2_{\perp}(H) \sin^2 \biggl(\frac{\omega_c z}{2 v_F} \biggl) \biggl] \cos \biggl[  \frac{v_x(p_y)}{v_F} k z \biggl]  ,
\end{eqnarray}
where $g$ is renormalized electron coupling constant,
$x_1-x=z v_x(p_y)/v_F$.
Here, we consider Eq.(18), taking into account that electron velocity
component along conducting ${\bf x}$ axis is
\begin{equation}
v_x(p_y) = v_F [1 + \alpha \cos(p_y b')] ,
\end{equation}
where $\alpha = \sqrt{2} t_b / t_a \simeq 0.14$ [23].
Under condition $\alpha \ll 1$, Eq.(18) can be simplified:
\begin{eqnarray}
\frac{1}{g}= \int^{\infty}_{\frac{v_F}{\Omega}} \frac{dz}{z}
&&\cos \biggl( \frac{2 \beta \mu_B H z}{v_F} \biggl) \cos(k z) [J_0(\alpha k z) - m J_2(\alpha kz)]
\nonumber\\
&&\times \biggl[1-2 l^2_{\perp}(H) \sin^2 \biggl(\frac{\omega_c z}{2 v_F} \biggl) \biggl] ,
\end{eqnarray}
where $m=1$ for d-wave-like nodal superconducting order parameter, whereas 
$m=0$ for d-wave-like node-less and s-wave ones.
It is important that, in the absence of the paramagnetic effects, Eq.(20) describes
the Reentrant superconductivity [19] with transition temperature being increasing 
function of a magnetic field.
Therefore, we call superconducting phase, described by Eq.(20), the hidden Reentrant
superconductivity.

We can simplify Eq.(20) by using the following relationship:
\begin{equation}
\frac{1}{g} = \int^{\infty}_{\frac{v_F}{\Omega}}
\frac{2 \pi T_c dz}{v_F \sinh ( \frac{2 \pi T_c z}{v_F} )} ,
\end{equation}
where $T_c$ is superconducting transition temperature in the absence
of a magnetic field.
As a result, we obtain
\begin{eqnarray}
&&\ln \biggl (\frac{H^{b'}_{c2}}{H^*} \biggl) = \int^{\infty}_{0} \frac{dz}{z} \cos 
\biggl( \frac{2 \beta \mu_B Hz}{v_F}
\biggl)
\nonumber\\
&&\times \biggl\{\cos(kz) [J_0(\alpha k z)- m J_2(\alpha k z)]
\nonumber\\
&&\times \biggl[1 - 2 l^2_{\perp}(H) \sin^2 \biggl( \frac{\omega_c z}{2 v_F} \biggl) \biggl] -1 \biggl\} ,
\end{eqnarray}
where $\mu_B H^* = \pi T_c / 2 \gamma$,
$\gamma$ is the Euler constant.
We numerically find maxima of $H^{b'}_{c2}(k)$ as a function of wave vector, $k$,
of the order parameter (17) for m=1 and m=0. 
We come to the conclusion that experimental value, $H^{b'}_{c2} \simeq 6 \ T$,
can be obtained for d-wave-like nodal order parameter at $\beta \simeq 0.85$,
 whereas for d-wave-like node-less and s-wave order parameters it corresponds 
$\beta \simeq 0.5$.
Note that the so-called $g$-factor in (TMTSF)$_2$ClO$_4$ conductor is very close 
to its standard value, g= 2 [1], which corresponds to $\beta=1$ in
Eq.(22).
Therefore, we consider d-wave-like nodal superconductivity to be much more 
consistent with the experimental data, where the calculated value $\beta \simeq 0.85$
is close to its expected value, $\beta=1$.
It is not exactly equal to $1$, perhaps, due to slightly deviations of superconductivity 
in (TMTSF)$_2$ClO$_4$ from weak 
coupling scenario.
On the other hand, we consider d-wave-like node-less and s-wave parings as very
unlikely due to very low calculated value of $\beta \simeq 0.5$. 

\section{Conclusion}
\label{3}
We have suggested explanation of high values of the upper critical magnetic fields,
experimentally observed in (TMTSF)$_2$ClO$_4$ conductor [13,14,16], using 
d-wave-like nodal scenario of 
superconductivity. 
On the other hand, we anticipate that, for explanation of very high values of the upper critical magnetic fields [33,34] in a sister compound, (TMTSF)$_2$PF$_6$, in mixed superconducting-antiferromagnetic phase [34,35], it is necessary to consider either triplet 
superconducting pairing [10,23,24] or the coexistence of triplet and singlet superconducting
order parameters [36].

We are thankful to N.N. Bagmet, S.E. Brown, and Y. Maeno for useful discussions.
This work was supported by the NSF under Grants No DMR-0705986 and
DMR-1104512.





\begin{thebibliography}{00}
\bibitem{Shima-1}
Note that d-wave-like node-less order parameter, suggested in Ref.[27], 
may exist in (TMTSF)$_2$ClO$_4$ conductor due to the called anion gap 
in its Q1D electron spectrum [1].
\end{thebibliography}
\end{document}